\documentclass[prb,twocolumn,showpacs,amsmath,amssymb]{revtex4}
\usepackage{graphicx}
\usepackage{dcolumn}
\usepackage{bm}

\begin{document}

\preprint{APS/123-QED}

\title{Interplay of frustration and magnetic field in 
the two-dimensional quantum antiferromagnet Cu(tn)Cl$_{2}$} 

\author{A. Orend\'{a}\v{c}ov\'{a}}\email{alzbeta.orendacova@upjs.sk}%
\affiliation{Center of Low Temperature Physics, Faculty of Science, P. J. \v{S}af\'{a}rik University, 
Park Angelinum 9, 041 54 Ko\v{s}ice, Slovakia}%
\author{E. \v{C}i\v{z}m\'{a}r}%
\affiliation{Center of Low Temperature Physics, Faculty of Science, P. J. \v{S}af\'{a}rik University, 
Park Angelinum 9, 041 54 Ko\v{s}ice, Slovakia}
\author{J. S. Xia}
\affiliation{Department of Physics and the National High Magnetic Field Laboratory, University of Florida, 
Gainesville, FL 32611-8440, USA}
\author{L. Yin}
\affiliation{Department of Physics and the National High Magnetic Field Laboratory, University of Florida, 
Gainesville, FL 32611-8440, USA}
\author{D. M. Pajerowski}
\affiliation{Department of Physics and the National High Magnetic Field Laboratory, University of Florida, 
Gainesville, FL 32611-8440, USA}
\author{L. Sedl\'{a}kov\'{a}}%
\affiliation{Center of Low Temperature Physics, Faculty of Science, P. J. \v{S}af\'{a}rik University, 
Park Angelinum 9, 041 54 Ko\v{s}ice, Slovakia}%
\author{J. Hanko}
\affiliation{Center of Low Temperature Physics, Faculty of Science, P. J. \v{S}af\'{a}rik University, 
Park Angelinum 9, 041 54 Ko\v{s}ice, Slovakia}
\author{V. Zele\v{n}\'{a}k}
\affiliation{Institute of Chemistry, Faculty of Sciences, P. J. \v{S}af\'{a}rik University, Moyzesova 16, 
041 54 Ko\v{s}ice, Slovakia}
\author{M. Kaj\v{n}akov\'{a}}
\affiliation{Center of Low Temperature Physics, Faculty of Science, P. J. \v{S}af\'{a}rik University, 
Park Angelinum 9, 041 54 Ko\v{s}ice, Slovakia}
\author{S. Zvyagin}
\affiliation{Hochfeld-Magnetlabor Dresden (HLD), Forschungszentrum Dresden-Rossendorf, D-01314 Dresden, 
Germany}
\author{M. Orend\'{a}\v{c}}
\affiliation{Center of Low Temperature Physics, Faculty of Science, P. J. \v{S}af\'{a}rik University, 
Park Angelinum 9, 041 54 Ko\v{s}ice, Slovakia}
\author{A. Feher}
\affiliation{Center of Low Temperature Physics, Faculty of Science, P. J. \v{S}af\'{a}rik University, 
Park Angelinum 9, 041 54 Ko\v{s}ice, Slovakia}
\author{J. Wosnitza}
\affiliation{Hochfeld-Magnetlabor Dresden (HLD), Forschungszentrum Dresden-Rossendorf, D-01314 Dresden, 
Germany}%
\author{M. W. Meisel}
\affiliation{Department of Physics and the National High Magnetic Field Laboratory, University of Florida, 
Gainesville, FL 32611-8440, USA}

\date{\today}

\begin{abstract}
Specific heat and ac magnetic susceptibility measurements, spanning low temperatures ($T \geq 40$~mK) 
and high magnetic fields ($B \leq 14$~T), have been performed on a two-dimensional (2D) antiferromagnet 
Cu(tn)Cl$_{2}$ (tn = C$_{3}$H$_{10}$N$_{2}$). The compound represents an $S = 1/2$ spatially anisotropic 
triangular magnet realized by a 
square lattice with nearest-neighbor ($J/k_{B} = 3$~K), frustrating next-nearest-neighbor 
($0 < J^{\prime}/J < 0.6$), and interlayer ($|J^{\prime \prime}/J| \approx 10^{-3}$) interactions.      
The absence of long-range magnetic order down to $T = $ 60 mK in $B = 0$ and the $T^{2}$ 
behavior of the specific heat for $T \leq 0.4$~K and $B \geq 0$ are considered 
evidence of high degree of 2D magnetic order.  In fields lower than the saturation field, 
$B_{\text{sat}} = 6.6$~T, 
a specific heat anomaly, appearing near 0.8~K, is ascribed to bound vortex-antivortex pairs stabilized by 
the applied magnetic 
field.  The resulting magnetic phase diagram is remarkably consistent with the one predicted for the 
ideal square lattice, except that $B_{\text{sat}}$ is shifted to values 
lower than expected. Potential explanations for this observation, as well as the possibility of a 
Berezinski-Kosterlitz-Thouless (BKT) phase transition in a spatially anisotropic 
triangular magnet with the N\'{e}el ground state, are discussed. 
\end{abstract}

\maketitle

\section{Introduction}
Two-dimensional quantum antiferromagnets have attracted a significant amount of theoretical and 
experimental attention due to the unconventional magnetic properties resulting from the interplay 
between quantum fluctuations and geometrical frustration.\cite{Kohno,Sachdev,Xu}  One example is the 
$S = 1/2$ spatially anisotropic triangular antiferromagnet, which can be treated as a square lattice 
with the nearest-neighbor (\emph{nn}) interaction $J$ and frustrating  
next-nearest-neighbor (\emph{nnn}) interaction $J^{\prime}$, Fig.~1.  Between the limiting values 
of $J^{\prime} = 0$ (ideal square lattice) and $J^{\prime}/J \gg 1$ (spin chain), 
several phases exist for varying ratios of $J^{\prime}/J$,\cite{Merino, Weihong, Manuel, Weng, Pardini} 
while the presence of an applied magnetic field provides an additional constraint. 
Considerable theoretical interest in the spin liquid phase ($J^{\prime}/J$ $>$ 1) 
in a magnetic field\cite{Veillette, Alicea,Starykh1} was triggered by experimental studies of 
Cs$_{2}$CuCl$_{4}$, a spatially anisotropic triangular magnet with 
$J^{\prime}/J \approx$ 3.\cite{Coldea,Radu,Tokiwa}  

Recently, Cu(tn)Cl$_{2}$ has been identified as a potential model system for the realization of the 
spatially anisotropic triangular lattice from the collinear N\'{e}el phase 
($J^{\prime}/J$ $<$ 0.6).\cite{Zelenak}  For Cu(tn)Cl$_2$ studied in $B = 0$, no evidence for 
long-range magnetic order was observed down to 60~mK, and the data suggested intralayer interaction 
strengths of $J/k_B = 3$~K and $0 < J^{\prime}/J < 0.6$, while the interlayer coupling is  
$|J^{\prime \prime}/J| \approx 10^{-3}$. These interactions are described by the Hamiltonian
\begin{equation}
\mathcal H \;=\;J\,\sum_{nn} \mathbf{S}_i \cdot \mathbf{S}_j
\;+\;J^{\prime}\,\sum_{nnn} \mathbf{S}_{i} \cdot \mathbf{S}_{j} +\;J^{\prime \prime}\,\sum_{i,k} 
\mathbf{S}_i \cdot \mathbf{S}_k \;,
\end{equation}
where $i,j$ label intralayer spins and $k$ labels the interlayer ones.
\begin{figure}[hb]
\includegraphics[width=2.85in]{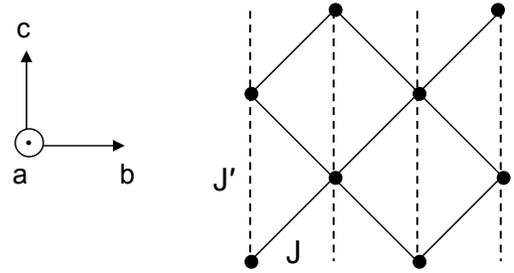}
\caption{\label{1} Realization of Heisenberg model of a spatially anisotropic triangular lattice within 
a single $bc$ layer in Cu(tn)Cl$_{2}$. The layers are stacked along the $a$ direction. The full circles denote Cu$^{2+}$ ions.}
\end{figure}

The motivation of the present work was to explore the response of Cu(tn)Cl$_2$ in $B \neq 0$, 
especially at low temperatures, $T \ll J/k_B$.  For this purpose, experimental 
specific heat and ac susceptibility studies were performed over a wide range of temperatures 
(40~mK $\leq T \leq 10$~K) and magnetic fields $(0 \leq B \leq 14$~T).  
The low value of the intralayer exchange coupling affords easy access to the 
magnetic phases below and above the saturation field.  On the basis of the magnetic field induced 
features observed in the specific heat and ac susceptibility, a magnetic phase diagram is  
constructed and analyzed within a model of a field induced Berezinski-Kosterlitz-Thouless (BKT) 
phase transition theoretically predicted for the pure 2D square lattice with zero \emph{nnn} 
coupling.\cite{Cuccoli}  It is noteworthy that the long-range orderings in some quasi-2D, 
square lattice systems, namely Sr$_2$CuO$_2$Cl$_2$ and several Cu(pyz)$_2$-based materials (pyz = pyrazine), 
have been placed close to the BKT transition.\cite{Sengupta, Xiao, Goddard}

Our presentation begins with a discussion of the sample and experimental details, which are followed 
by a description of our experimental results.  Next, an analysis and a discussion of the results are  
given before the salient points are assembled into the magnetic phase diagram.  The paper concludes 
with a summary and some comments about possible future directions. 

\section{Sample and Experimental details}
The crystal structure of Cu(tn)Cl$_{2}$ (tn = C$_{3}$H$_{10}$N$_{2}$), established at 150~K, is 
orthorhombic (space group Pna2$_{1}$) with the lattice parameters $a$ = 17.956~\AA, $b$ = 6.859~\AA, 
and $c$ = 5.710~\AA.\cite{Zelenak}  The structure consists of covalently bonded ladders running along the 
$c$-axis, while the adjacent ladders in the $bc$-plane are linked through intermolecular 
N$-$H$\cdots$Cl hydrogen bonds formed by all four H atoms of the amino groups. In the $a$ direction, 
the layers are connected by weak C$-$H$\cdots$Cl type interactions.  The strongly elongated octahedra 
coordinating the Cu(II) ions stabilize the $d_{z^{2}}$ electronic ground state. 
Consequently, the propagation of exchange pathways between Cu(II) ions leads to the formation of a 
spatially anisotropic triangular lattice in the $bc$-plane (Fig.~1).

The synthesis of all Cu(tn)Cl$_2$ samples followed 
the established procedure,\cite{Zelenak} which produces polycrystals that are powdered 
and pressed into pellets (nominally 3~mg to 10~mg) for the specific heat studies or placed 
into appropriate specimen holders for the magnetic susceptibility investigations. 
A sample with a mass of 26~mg was used in the ac magnetic measurements.  

Using several experimental probes and instruments, the specific heat measurements were performed 
over the temperature range from 100~mK to 10~K and in magnetic fields up to 14~T.  More specifically, 
the studies in the millikelvin temperature region and in magnetic fields up to 2.5~T were performed 
using a relaxation calorimeter\cite{Riegel} mounted on a dilution refrigerator. 
A commercial (Quantum Design PPMS) 
device was used for the specific heat studies over the temperature range from 1.8~K to 10~K and in fields 
up to 9~T, while the low temperature studies down to 0.35~K and in fields up to 14~T utilized another 
commercial instrument equipped with a $^{3}$He insert. In each instance, the contribution of the 
background addenda was determined in separate runs.  Finally, the separate study of the specific heat 
of a diamagnetic isomorph, Zn(tn)Cl$_2$, allowed the phonon contribution to be determined. 

The magnetic susceptibility studies were performed with ac (232~Hz) mutual inductance coils mounted on a 
dilution refrigerator equipped with a 10~T magnet. With the sample immersed in pure $^3$He that 
provided intimate thermal contact with the mixing chamber, the in-phase and out-of-phase 
signals of the susceptibility were recorded by a two channel lock-in amplifier. 
Typically the data were obtained by isothermal field sweeps at a rate of 50~mT/min, and the data were 
independent of the direction of the field sweep.   

\section{Experimental results}

\subsection{Specific heat in \boldmath{$B \neq 0$}}

\begin{figure}[b]
\includegraphics[width=3.25in]{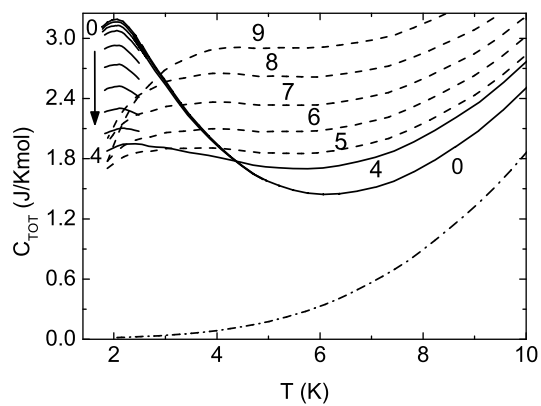}
\caption{\label{2} Temperature dependence of the total specific heat, C$_{\mathrm{TOT}}$, 
of Cu(tn)Cl$_{2}$ in various magnetic fields.  
Solid lines represent data for $B$ = 0, 0.5, 0.75, 1, 1.5, 2, 2.5, 3, 3.5 and 4 T, while the 
dashed lines correspond to $B$ = 5, 6, 7, 8 and 9 T. The dot-dashed line represents the 
specific heat of the diamagnetic isomorph Zn(tn)Cl$_{2}$.}
\end{figure}

Evidence of the presence of 2D magnetic correlations was initially observed as a round maximum in 
the temperature dependence of the total specific heat, $C_{\mathrm{TOT}}(T)$, near 2~K in 
$B = 0$,\cite{Zelenak} and our present work found this feature to evolve with magnetic fields up 
to 9~T, Fig.~2.  In fact, the influence of the magnetic field can be separated into two regimes. 
For $B < 4$~T, the height of the specific heat maximum, $C_{\mathrm{max}}$, decreases 
with increasing field while its 
temperature, $T_{\mathrm{max}}$, remains nearly unchanged, whereas  
for $B > 4$ T, $C_{\mathrm{max}}$ increases with increasing field 
and $T_{\mathrm{max}}$ shifts towards higher temperatures.  The phonon contribution to the 
specific heat can be approximated by the specific heat of the diamagnetic 
isomorph Zn(tn)Cl$_{2}$, Fig.~2.
These high temperature measurements were extended to lower temperatures in $B \leq 2.5$~T, and the 
results of the total specific heat are shown in Fig.~3.  
These studies revealed the presence of an anomaly appearing at about 0.8 K when $B \neq 0$, and this 
feature develops with increasing magnetic field (inset of Fig.~3).  
The magnetic field and temperature dependences of this anomaly were measured in $B \leq 14$~T, 
Fig.~4, where this feature reaches a maximum in 4~T at 0.8~K and decreases in magnitude, with a 
simultaneous shift to lower temperatures, until its presence is no longer resolved in fields greater 
than 7~T.

\begin{figure}[b]
\includegraphics[width=3.25in]{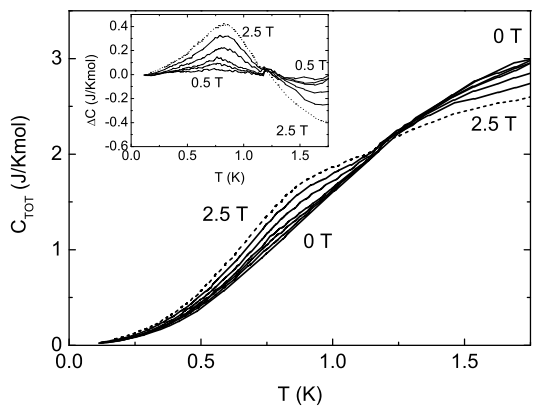}
\caption{\label{3} Temperature dependence of the total specific heat of 
Cu(tn)Cl$_{2}$ in $B$ = 0, 0.5, 0.75, 1, 1.5, 2 and 2.5 T.  
Inset: Temperature dependence of the difference $\Delta C = [C(T,B)-C(T,0)]$ between the total specific heat 
in finite and zero magnetic fields.}
\end{figure}

\subsection{Magnetic susceptibility in \boldmath{$B \leq 10$} T}

The results of isothermal ac susceptibility studies in $B \leq 10$~T 
are shown in Fig.~5, and no significant hysteresis was observed between the up and down sweeps.  
The data do not possess any sharp anomalies corresponding to a phase transition, but the appearance 
of a shoulder near 6~T corresponds 
to the field where the low temperature specific heat anomaly vanishes. Unlike the low temperature 
specific heat anomaly, the shoulder survives up to 1~K, suggesting that it is associated 
with the saturation magnetic field $B_{\mathrm{sat}}$.

\begin{figure}
\includegraphics[width=3.25in]{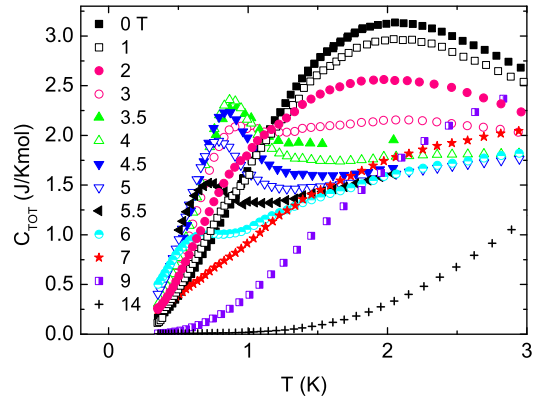}
\caption{\label{4} (Color online) Temperature dependence of the total specific heat of
Cu(tn)Cl$_{2}$ in various magnetic fields.}
\end{figure}
\begin{figure}
\includegraphics[width=3.25in]{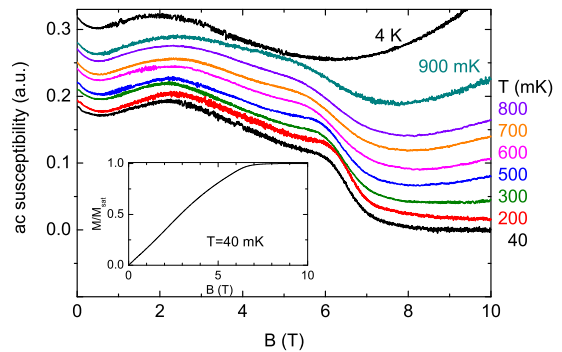}
\caption{\label{5} (Color online) Magnetic field dependence of isothermal ac susceptibility of 
Cu(tn)Cl$_{2}$ at various temperatures.  
Inset: Magnetic field dependence of the normalized isothermal magnetization at 40~mK.}
\end{figure}

\section{Analysis and Discussion}

\subsection{Magnetic correlations in \boldmath{$B = 0$}}

As stated earlier, previous specific heat studies in $B = 0$ did not indicate 
any magnetic phase transition down to 60~mK, and a clear quadratic dependence was observed 
at low temperatures. The latter coincides with the expected 2D character of short-range magnetic 
correlations,\cite{Zelenak} since the spin wave analysis of low dimensional models with \emph{nn} 
interactions predicts a $T^{2}$ behavior 
for the low temperature specific heat of square and triangular lattices and a $T$ dependence for a 
linear chain model.\cite{Bernu} The absence of a  $\lambda$-like anomaly associated with 
magnetic ordering is another intriguing feature of the specific heat data. 

In the absence of the frustration ($J^{\prime}$ = 0), the ordering temperature 
$T_{\text{N}}$ of the isotropic square lattice with interlayer coupling $J^{\prime \prime}$ 
can be expressed as\cite{Nelson,Chubukov}
\begin{equation}
k_{B}T_{\text{N}}\;=\;J^{\prime \prime}\left(\frac{M}{M_{0}}\right)^{2} \left(\frac{\xi}{a}\right)^{2}\;,
\label{temp}
\end{equation}
where $M/M_{0}$ is a staggered magnetization, $a$ represents a lattice constant, and $\xi$  
is an intralayer correlation length given by 
\begin{equation}
\frac{\xi}{a}\;=\;0.5\; \frac{\exp \left({\frac{2\pi \rho_{s}}{k_B T}}\right)}
{1+\left(\frac{k_B T}{2\pi \rho_{s}}\right)}\;\;\;.
\label{cor}
\end{equation}
Here, $\rho_{s}$ is a spin stiffness, and for the isotropic square lattice with intralayer \emph{nn} 
exchange coupling $J$, it can be expressed as $\rho_{s} \cong 0.18 J$. Using the parameters 
$|J^{\prime \prime}/k_B | \approx 3$~mK, $|J/k_B | \approx 3$~K, determined in Ref. \onlinecite{Zelenak}, and $(M/M_{0})^{2} \approx  0.3$, 
a phase transition in Cu(tn)Cl$_{2}$ might be expected at $T_{\text{N}} \approx 0.8$~K. 
The absence of the phase transition down to 60~mK   
may suggest a significant reduction of both the spin stiffness and the staggered magnetization 
as a consequence of frustrating \emph{nnn} $J^{\prime}$ coupling.\cite{Merino,Weihong,Manuel,Nelson,Chubukov} 
Apart from the aforementioned reduction of staggered magnetization, 
the weakness of the interlayer coupling can also lead to the absence of a detectable phase transition at 
finite temperatures in specific heat measurements. 
Recent Monte Carlo studies of the specific heat of an isotropic square lattice with various 
interlayer couplings revealed that, for $J^{\prime \prime}/J \alt 0.015$, the peak 
associated with the phase transition vanishes.\cite{Singh}  Since $|J^{\prime \prime}/J| \approx 10^{-3}$ for 
Cu(tn)Cl$_2$, the ordering effects might not be observed in the specific heat even in the absence of 
frustration.  While experimental studies of Cu(pyz)$_{2}$(ClO$_{4}$)$_{2}$ are consistent 
with this scenario,\cite{Lancaster} the observation of a phase transition in the specific heat of 
Cu(H$_{2}$O)$_{2}$(C$_{2}$H$_{8}$N$_{2}$)SO$_{4}$ with a comparable 
$J^{\prime \prime}/J$ ratio shows the need to consider additional effects.\cite{Kajnak}

\subsection{Magnetic correlations in \boldmath{$B \neq 0$}}

\subsubsection{Specific heat for $T \geq J/k_B$}

After the subtraction of the phonon contribution,  
approximated by the specific heat of the diamagnetic isomorph Zn(tn)Cl$_{2}$ (Fig.~2),  
two different regimes can be distinguished in the  behavior of $C_{\mathrm{max}}$ for $T > 1.8$~K.  
For $B \lesssim 4$~T, $C_{\mathrm{max}}$ decreases with increasing field  
while $T_{\mathrm{max}}$ remains nearly unchanged, whereas for $B \gtrsim 5$~T, 
$C_{\mathrm{max}}$ increases with increasing field and $T_{\mathrm{max}}$ shifts towards 
higher temperatures, Fig.~6. Such a qualitative behavior has been predicted for a linear (1D) 
chain with $nn$ coupling ${\cal J}$,\cite{Bonner, Klumper} where the crossover between the two regimes occurs at a saturation field 
\begin{equation}
g\mu_{\text{B}}B_{\text{sat}} = 2{\cal J}\;.
\end{equation}
Using $g = 2.12$ and ${\cal J}/k_B = 4$~K, which correspond to the best estimates of the parameters 
resulting from the linear chain model as applied to Cu(tn)Cl$_2$ in Ref.~\onlinecite{Zelenak}, 
Eq.~4 predicts $B_{\text{sat}} = 5.6$~T, a value that coincides rather well with the crossover 
region appearing between 4~T and 6~T. However, 
in comparison with the linear chain, which is characterized by a decrease of $T_{\text{max}}$ 
with increasing field for fields lower than $B_{\text{sat}}$ and an increase for higher fields, 
the observed shift of $T_{\text{max}}$ with respect magnetic field behaves differently.  
Alternatively, the effect of an external magnetic field on the short-range correlations and the 
thermodynamic properties of a square lattice have been theoretically investigated,\cite{Cuccoli}  
and the saturation field is 
\begin{equation}
g\mu_{\text{B}}B_{\text{sat}} = 4J\;.
\end{equation}
For Cu(tn)Cl$_{2}$, this model predicts $B_{\text{sat}} = 8.4$ T, which is too high to correspond 
to the observed crossover region.  

\begin{figure}
\includegraphics[width=3.25in]{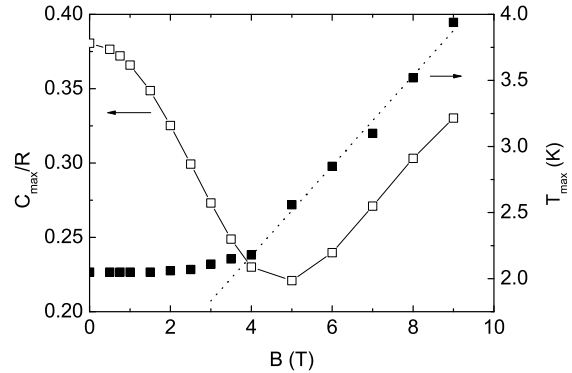}
\caption{\label{6} Magnetic field dependence of the value for the maximum of the 
Cu(tn)Cl$_{2}$ magnetic specific heat, $C_{\mathrm{max}}$, divided by the gas constant $R$ (open squares). 
The solid line is a guide for eyes. 
Magnetic field dependence of the temperature, $T_{\mathrm{max}}$, denoting the position of $C_{\mathrm{max}}$ (full squares). 
The broken line is a linear fit (see text).}
\end{figure}

This qualitative comparison of the limiting 1D and 2D theoretical models with the experimental data 
suggests the existence of two phases in Cu(tn)Cl$_{2}$, namely a low field phase stable 
in the magnetic fields below the crossover region where antiferromagnetic correlations 
prevail and a high field paramagnetic phase stabilized in the fields above the crossover region. 
However, the atypical magnetic field dependence of $T_{\mathrm{max}}$ suggests a more complex low field phase.  
Indeed, a closer look at the behavior shown in the Fig.~6 reveals that, 
within experimental uncertainty, $T_{\text{max}}$ remains constant for $B < 2$~T, 
before increasing slightly in the fields from 
2~T to 4~T, and then experiencing a quasi-linear increase for the fields above 4~T, as demonstrated by 
a linear fit performed in the range from 4~T to 9~T. 

\subsubsection{Specific heat for $T \leq J/k_B$}

The quantitative analysis of magnetic specific heat, $C_{M}$, at temperatures $T < 0.4$~K and 
$B < 2.5$ T reveals the $T^{2}$ dependence, as can be seen in Fig.~7, where the data are 
plotted as $C_{M}T^{2}$ vs.~$T^{4}$.  This approach assumes the magnetic specific heat can be expressed 
as a sum of $a/T^{2}$ and $bT^{2}$ terms, where the first term corresponds to the 
contribution of the nuclear spins and/or long-range correlations between electronic spins, 
and the second term is associated with 2D short-range correlations.  A set of  linear 
fits of the individual $C_{M}T^{2}$ vs.~$T^{4}$ dependences for fixed magnetic field was performed 
in the temperature range from nominally 150~mK to 400~mK. The fitting procedure revealed a monotonic 
rise of the coefficient $b$ (in units of J/(K$^{3}$ mol)) with increasing magnetic field and 
$a \approx 0$. The monotonic increase can be approximated by a linear dependence   
\begin{equation}
b(B)\;=\;1.36\;+\;0.30\,B \;,
\end{equation}
as shown in the inset of Fig.~7.  
The observed low temperature $T^{2}$ behavior of $C_{M}(T)$ suggests the preservation 
of the mainly 2D character of the magnetic correlations in $B \neq 0$. Furthermore, 
the field does not introduce an energy gap to the excitation spectrum, which remains 
gapless as expected for the square lattice in magnetic fields below the saturation value.\cite{Cuccoli} 

\begin{figure}
\includegraphics[width=3.25in]{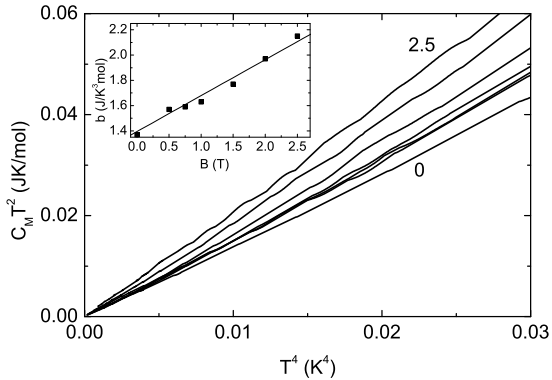}
\caption{\label{7} $C_{\mathrm{M}}T^{2}$ vs.~$T^{4}$ dependence for Cu(tn)Cl$_2$ in 
$B$ = 0, 0.5, 0.75, 1, 1.5, 2 and 2.5 T. 
Inset: Magnetic field dependence of the coefficient $b$, Eq.~6. The solid line represents a linear fit.}
\end{figure}

\begin{figure}
\includegraphics[width=3.25in]{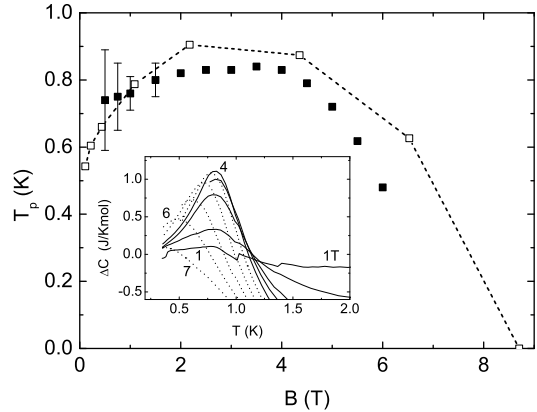}
\caption{\label{8}  Magnetic field dependence of the peak position, $T_{\text{p}}$, of the low temperature  
specific heat anomalies (full squares) in Cu(tn)Cl$_2$. The theoretical prediction for the field induced BKT phase transition 
for the isotropic square lattice is denoted by the open squares and the broken line. 
Inset: Temperature dependence of the difference between the specific heat in finite  
and zero magnetic field. The solid lines represent $B$ = 1, 2, 3, 3.5 and 4 T, and 
the dashed lines correspond to $B$ = 4.5, 5, 5.5, 6 and 7 T.}
\end{figure}

To more precisely trace the position of the low temperature peak, $T_{\text{p}}$, appearing at about 
0.8~K, as a function of magnetic field, the specific heat in zero magnetic field was chosen as a 
reference background that was subtracted from the specific heat in nonzero magnetic field, inset of Fig.~8.    
Using the low-field data obtained in the same way (inset of Fig.~3), the magnetic field dependence of 
$T_{\text{p}}$ can be extracted (Fig.~8).  It can be seen that a naive extrapolation to zero field 
provides $T_{\text{p}}\approx 0.7$~K, which is close to the value of the transition temperature 
$T_{\text{N}} \approx 0.8$~K estimated for the isotropic square lattice. This coincidence might support 
the suggestion that the weakness of the interlayer coupling prevents the observation of the phase 
transition in the specific heat.  An extrapolation of the high-field experimental data yields 
$B(T_{\mathrm{p}} \rightarrow 0) \approx 6 - 7$~T, which is lower than $B_{\mathrm{sat}} = 8.4$~T 
calculated for the isotropic square lattice.

Finally, to determine $B_{\mathrm{sat}}$ from the specific heat data, 
we analyzed the $C_{M}(T)$ data in 9~T and 14~T.  While the peak height of $C_{M}(T)$ in 9~T is 
still much lower than $C_{\mathrm{max}}/R = 0.438$, the theoretical prediction for the 
$S = 1/2$ ideal paramagnet (Fig.~6), the corresponding $B = 14$~T value of $C_{\mathrm{max}}/R = 0.42$ is 
close to the expected result. 
For $B > B_{\text{sat}}$, a gap $\Delta$ opens in the spin excitation spectrum,  
developing linearly with magnetic field as\cite{Radu}
\begin{equation}
\Delta = g\mu_{\text{B}}(B - B_{\text{sat}})\;.
\label{gap}
\end{equation}  
In the temperature region where 
thermal fluctuations overcome the interlayer coupling and $T < \Delta/ k_B $, 
a 2D character of magnon spectra 
can be expected, resulting in\cite{Radu}
\begin{equation}  
C_{M}(T) \; \approx \; \frac{1}{T} \; \exp(-\Delta/k_B T) \;. 
\end{equation}
Fitting the $C_M(T < 1$ K, $B = 9$~T) and $C_M(T < 3$ K, $B = 14$~T) data yields 
$\Delta/k_B = 3.4$~K and 10.5~K, respectively, 
and it follows from Eq.~\ref{gap} that $B_{\text{sat}} = 6.6 \pm 0.1$~T, when assuming $g = 2.12$.  
It is important to note that 
the significant differences between the observed $B_{\mathrm{sat}} = 6.6$~T and the 8.4~T value 
expected for the isotropic square lattice indicate that  
Cu(tn)Cl$_{2}$ is indeed affected by the presence of frustrating \emph{nnn} interactions, 
since the interlayer coupling is expected to increase $B_{\mathrm{sat}}$ as\cite{Sengupta} 
\begin{equation}
g\mu_{\text{B}}B_{\text{sat}} \;= \; 4J \;+ \; 2J^{\prime \prime} \;.
\label{Bsat}
\end{equation}
     
\subsubsection{AC susceptibility for $T \leq J/k_B$}

The ac susceptibility data, $\chi(T,B) = \partial M/\partial B$, 
can be integrated to obtain the magnetization 
$M(T,B)$, as shown in the inset of Fig.~5, and $M(T\rightarrow 0,B)$ data is commonly 
used to establish $B_{\mathrm{sat}}$, if the saturation plateau is accessible.  However, 
in many instances, the transition to the fully-polarized state is broadened by several 
effects, including issues related to finite temperature, orientation of the microcrystals, and 
finite size effects, so an extrapolation is used to establish a value for $B_{\mathrm{sat}}$. 
Our ac susceptibility data shown 
in Fig.~5 suggest $B_{\mathrm{sat}} = 6.5 \pm 0.2$~T for $T \leq 200$~mK (when considering 
$B_{\mathrm{sat}}$ as the mid-point of the region where $\chi(B)$ is most strongly changing), 
and this value agrees with the one extracted from the analysis of the $C_M$ data at 9~T and 14~T.  

The crossover region, $\Delta B_{\text{sat}}$, that spans from the region from about 
6~T to 7~T at low temperature, Fig.~5, merits further analysis.  For example, 
when $k_{B}T \approx 0.1J$, thermal smearing of the crossover is not expected.  
Alternatively, the powder character of  the sample might induce a scatter of $B_{\text{sat}}$ 
that can be evaluated as $B_{\text{sat}}^{\parallel}-B_{\text{sat}}^{\perp}$. 
Previous electron paramagnetic resonance studies provided values for 
the anisotropic g-tensor of $g_{\parallel} = 2.25$ and $g_{\perp} = 2.05$.\cite{Zelenak}  
Assuming an ideal square lattice and taking $J$/$k_{B}$ = 3 K, then
\begin{equation} 
B_{\text{sat}}^{\parallel,\perp} \; = \;
\frac{4 J}{g_{_{\parallel,\perp}} \; \mu_{\text{B}}}\;\;\;,
\end{equation}
yields 
$\mid\Delta B_{\text{sat}}\mid \leq 0.8$~T, which is less than but similar in magnitude to 
the observed width of the crossover.  In addition, the finite size of the particles  
comprising the powder-like sample might cause broadening by limiting the in-plane 
correlation length $\xi$ at the lowest temperatures.  
Perhaps more fundamentally, the spatial extent of $\xi$ near $B_{sat}$ 
might be restricted by the underlying magnetic frustration, which suppresses the  
antiferromagnetic correlations.

\subsection{Magnetic phase diagram of Cu(tn)Cl\boldmath{$_{2}$}}

Prior to building the magnetic phase diagram, the correlations between the shift of $T_{\mathrm{p}}$ 
and the position of high temperature specific heat maximum $T_{\mathrm{max}}$ with respect 
to magnetic field are worth noting (Figs.~6 and 8). The monotonic increase of  $T_{\mathrm{p}}$ with 
increasing magnetic field 
up to about 2 T corresponds to the relative field independence of $T_{\mathrm{max}}$ in this 
field region. The relative field independence of $T_{\mathrm{p}}$ in the field from 2 to 4 T 
corresponds to the slight increase of $T_{\mathrm{max}}$, and finally, the rather rapid 
decrease of $T_{\mathrm{p}}$ above 4 T corresponds to a linear increase of $T_{\mathrm{max}}$. 

This relationship between the low temperature anomaly and the high temperature maximum suggests that the 
magnetic field mainly affects the short-range correlations. This interpretation is supported by the monotonic 
increase of $T_{\mathrm{p}}$ observed in low fields, and this behavior is typical for 
the Berezinski-Kosterlitz-Thouless (BKT) transition theoretically predicted in the 
classical limit for low dimensional magnets in uniform magnetic fields.\cite{Villain, Pires}      
Quantum Monte Carlo studies of the $S = 1/2$ isotropic square lattice in a magnetic field also 
revealed field-induced XY behavior, and a BKT transition at a finite temperature, 
$T_{\text{BKT}}$, was identified for sufficiently strong magnetic fields.\cite{Cuccoli}  
Neither the presence of spatial anisotropy nor the introduction of 
the frustrated $nnn$ coupling are expected to qualitatively change the physical picture 
derived for the isotropic square lattice.\cite{private}  Consequently, the measured  
$T_{\text{p}}$ vs.~$B$ dependence can be compared to the theoretical 
prediction of $T_{\text{BKT}}$ vs.~$B$ calculated for $g$ = 2.12 and $J/k_{\text{B}} = 3$ K, and  
the comparison yields surprisingly good agreement between the theoretical predictions and data (Fig.~8). 
It should be noted that the authors of Ref.~\onlinecite{Cuccoli} 
identified the BKT phase transitions at $T_{\text{BKT}}$, which lies below $T_{\text{p}}$. 
However, the determination of $T_{\text{BKT}}$ directly from the experiment is not as straightforward 
as the determination of $T_{\text{p}}$. 
From this point of view, one must be aware that the experimental determinations of the potential  
$T_{\text{BKT}}$ are overestimated.

Recent quantum Monte Carlo calculations of a $B$ vs.~$T$ magnetic 
phase diagram performed for the system of a  tetragonal lattice with intralayer coupling $J$ 
and interlayer coupling $J''$ revealed that, for sufficiently large spatial anisotropy $J/J''$, 
the system preserves nonmonotonic $B$ vs.~$T$ behavior typical for the BKT transition on the ideal 
square lattice.\cite{Sengupta}  The enhancement of the critical temperatures and the saturation field 
with respect to the ideal 2D case is another effect of the interlayer coupling.  The predictions were 
compared with the experimentally established phase diagram of  
[Cu(HF$_{2}$)(pyz)$_{2}$]BF$_{4}$, a representative of the quasi-2D Heisenberg square lattice with a 
finite temperature phase transition to 3D long-range order in zero magnetic field, and excellent 
agreement was found.\cite{Sengupta}  Unlike the phase diagram of 
Cu(tn)Cl$_{2}$, the experimental data of [Cu(HF$_{2}$)(pyz)$_{2}$]BF$_{4}$ are significantly 
shifted above the theoretical prediction for the ideal square lattice model.  The fact that our data 
lie significantly lower, i.e.~below the theoretical predictions for the BKT transition (Fig.~8), 
supports the conjecture about the important presence of frustration.

\begin{figure}
\includegraphics[width=3.25in]{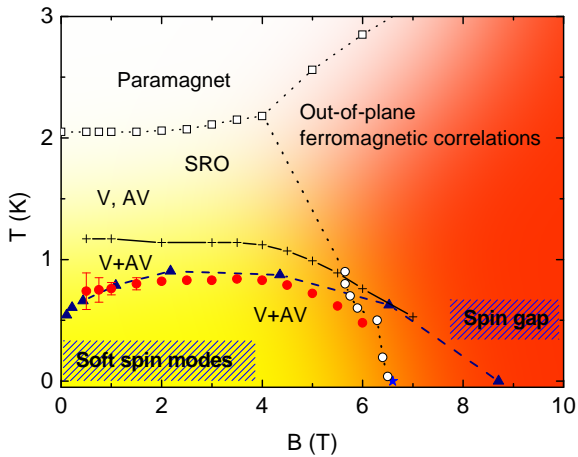}
\caption{\label{phase diagram} (Color online) 
Extended magnetic phase diagram of Cu(tn)Cl$_{2}$. The open squares represent 
the $T_{\mathrm{max}}$ vs.~$B$ dependence, while the pluses (+) correspond to the line constructed from the 
condition that $\delta_{T} S(T,B)- \delta_{T} S(T,0)=0$, see text. The theoretically predicted BKT transition 
and the experimental $T_{\mathrm{p}}$ vs.~$B$ dependence are denoted by full triangles and full circles, respectively. 
Open circles correspond to the positions of the mid-points in the shoulders of ac susceptibility data. 
The star represents the $B_{\text{sat}}$ as extracted from an analysis of the specific heat data measured 
in 9~T and 14~T. The dotted and solid lines are guides for eyes. 
The regions of short-range order (SRO) and free or bound vortex (V) and antivortex (AV) pairs are identified.  
Detailed descriptions of each $B-T$ region are given in the text.}
\end{figure}

Considering all the ac susceptibility and specific heat features together, we can identify the 
major regions in the extended magnetic phase diagram for Cu(tn)Cl$_{2}$ as a triangular magnet 
from the N\'{e}el phase (Fig.~\ref{phase diagram}).  At temperatures above 2 K, the 
system behaves as a paramagnet in all magnetic fields. The paramagnetic region is separated by 
the line determined by the $T_{\mathrm{max}}$ vs.~$B$ dependence.  At temperatures below the line, 
2D antiferromagnetic correlations develop, forming free vortices (V) and antivortices (AV) in the $xy$ plane. 
The vortices are stabilized by the magnetic field, whose direction defines the $z$-axis in the spin space.  
The formation of bound V-AV pairs begins at temperatures below a line derived from the requirement that the 
difference of the entropy derivatives $[\delta_{T} S(T,B) - \delta_{T} S(T,0)]$ is zero.\cite{Cuccoli}  
This quantity equals the difference of the specific heat $[C(T,B)-C(T,0)]$ divided by temperature $T$. 
The intensity of the pairing process culminates at the temperatures defined by the position of the 
low temperature specific heat anomaly.  As was shown in the theoretical studies of the BKT 
transition,\cite{Cuccoli, Sengupta} the temperature of the BKT transition itself is about $20-30$\% 
lower than the position of the specific heat maximum. Consequently for Cu(tn)Cl$_{2}$, 
the potential BKT transition can be expected at temperatures lower than the critical line constructed 
from the $T_{\mathrm{p}}$ vs.~$B$ data. 

Above the critical magnetic field region determined from the position of the high field shoulder in the 
isothermal ac susceptibility data, out-of-plane ferromagnetic correlations induced by the magnetic field 
are stabilized, while in-plane spin correlations show paramagnetic behavior.\cite{thesis}  In this region, 
a spin gap appears in the excitation spectrum, and this gap was detected by the exponential character of the 
low temperature specific heat measured in 9~T. As shown in the phase diagram, the induced ferromagnetic 
correlations survive up to the temperatures on the critical line $T_{\mathrm{max}}$ vs.~$B$.  
Coincidence with the theoretical predictions for the square lattice,\cite{thesis} formation of the 
out-of-plane ferromagnetic correlations already begins at fields above 4~T as indicated  by the 
linear dependence of $T_{\mathrm{max}}$ vs.~$B$ above 4 T. 

The onset of quasi-long-range order (QLRO), induced by the field due to the stabilization of bound 
V-AV pairs expected for the unfrustrated square lattice at sufficiently low temperatures, 
should be reflected by the smaller entropy increase in finite magnetic field than in zero field.  
In other words, the presence of the V-AV pairs should lead to negative values of $[C(T,B)-C(T,0)]$ 
at low temperatures.  However, as can be seen from the insets of Figs.~3 and 8, in Cu(tn)Cl$_{2}$, 
the difference remains positive down to the lowest temperatures. The increasing specific heat with magnetic field 
can be ascribed to the combined effect of the interlayer coupling and the  
existence of soft spin modes, likely associated with the frustrated $nnn$ coupling, which can  
prevent the system from achieving full QLRO.

\section{Summary}

The response of Cu(tn)Cl$_{2}$ to an externally applied magnetic field has been investigated by 
specific heat and ac susceptibility studies at low temperatures, $T \geq 40$~mK, 
and in magnetic fields up to 14 T. 
Specific features induced by the magnetic field in the behavior of both quantities allowed 
a magnetic phase diagram, whose regions are rooted in the presence of a high degree of short-range order, 
to be constructed.  The quadratic temperature dependence of the specific heat observed below 0.4~K and in  
magnetic fields up to, at least, 2.5~T arises from the predominantly 2D character of magnetic correlations and 
identifies a gapless excitation spectrum as predicted for the isotropic square lattice. 
Furthermore, in finite magnetic fields up to 6~T, a field induced anomaly appears near 0.8~K, and 
this feature is associated with the existence of a Berezinski-Kosterlitz-Thouless (BKT) transition theoretically 
predicted for the isotropic square lattice.\cite{Cuccoli}  The dimensionality of the observed 
magnetic field induced transition is unresolved and will be the focus of future neutron and/or muon 
scattering studies.  The absence of a phase transition in zero magnetic field is interpreted as a 
consequence of a combined effect of the weak interlayer coupling and the frustration within the 
magnetic layers.  The saturation magnetic field value, as extracted from the specific heat and 
ac susceptibility data, deviates significantly from the one predicted for an isotropic square 
lattice, and this result is conjectured to be a consequence of the frustrating magnetic interactions, 
in the $bc$ layer, that have have been confirmed recently by quantum mechanical calculations.\cite{Pavarini} 

Finally, it is noteworthy that a $T^{2}$ dependence of the specific heat has been reported for 
several frustrated 2D systems in zero magnetic field and was ascribed to various scenarios.  
While only a weak field dependence of the low temperature specific heat has been observed in  
Kagom\'{e}\cite{Ramirez} and triangular\cite{Nakatsuji} compounds, the linear increase of the 
$b$ coefficient, as observed in Cu(tn)Cl$_{2}$, points to the fact that even an infinitesimal field 
is able to introduce the changes. This sensitivity also supports the applicability of the BKT model to  
Cu(tn)Cl$_{2}$; however, only to some extent. In the spin wave region, the magnetic field should 
decrease the specific heat and, correspondingly, the $b$ coefficient, as expected for the square lattice. 
Since the opposite tendency was observed, our experimental data suggest the presence of soft spin modes 
that are possibly connected with frustration and interlayer coupling.  A spin wave analysis of the triangular magnet from the N\'{e}el 
phase in a magnetic field would be desirable to elucidate these observed differences.  
Recently, the BKT description has been successfully used\cite{Chern} to explain the 2D spin freezing transition 
observed in NiGa$_{2}$S$_{4}$, a model system for the $S$ = 1 isotropic triangular Heisenberg lattice.\cite{Mac}

In conclusion, the analysis of experimental data suggests that the $S = 1/2$ spatially anisotropic 
triangular magnet from the collinear N\'{e}el phase undergoes a Berezinski-Kosterlitz-Thouless transition 
induced by an applied magnetic field.  Theoretical studies of the collinear N\'{e}el phase in a magnetic field 
are necessary to specify the proper range of $J$, $J^{\prime}$  parameters where the BKT transition can occur.  
Microscopic magnetic studies of Cu(tn)Cl$_{2}$, possibly employing neutron and/or muon scattering techniques, are needed to 
clarify the nature of the ground state in zero and nonzero magnetic fields.    

\begin{acknowledgments}

One of us (A.~O.) has benefited from many fruitful discussions with Tommasso Roscilde and Eva Pavarini. 
We are grateful Sasha Chernyshev for the discussions about low temperature specific heat in a magnetic field.  
This work was supported, in part, by VEGA grant 1/0078/09, project APVV-0006-07, 
ESF RNP program ``Highly Frustrated Magnetism", NSF DMR-0701400, the NHMFL via cooperative agreement 
NSF DMR-0654118 and the State of Florida, Deutsche Physikalische Gesellschaft (DPG), and
EuroMagNET II.  Material support from US Steel Ko\v{s}ice s.r.o. is greatly acknowledged.
\end{acknowledgments}


\begin{thebibliography}{35}
\bibitem{Kohno} M.~Kohno, O.~A.~Starykh, and L.~Balents, Nat. Phys. {\bf 3}, 790 (2007).
\bibitem{Sachdev} S.~Sachdev, Nat.~Phys.~{\bf 4}, 173 (2008). 
\bibitem{Xu} C.~Xu and S.~Sachdev, Phys.~Rev.~B {\bf 79}, 064405 (2009).
\bibitem{Merino} J.~Merino, R.~H.~McKenzie, J.~B.~Marston, and C.~H.~Chung, J.~Phys.:~Condens.~Matter {\bf 11}, 2965 (1999).
\bibitem{Weihong} Z.~Weihong, R.~H.~McKenzie, and R.~P.~Singh, Phys.~Rev.~B {\bf 59}, 14367 (1999).
\bibitem{Manuel} L.~O.~Manuel and H.~A.~Ceccatto, Phys.~Rev.~B {\bf 60}, 9489 (1999).
\bibitem{Weng} M.~Q.~Weng, D.~N.~Sheng, Z.~Y.~Weng, and R.~J.~Bursill, Phys.~Rev.~B {\bf 74}, 012407 (2006).
\bibitem{Pardini} T.~Pardini and R.~R.~P.~Singh, Phys.~Rev.~B {\bf 77}, 214433 (2008).
\bibitem{Veillette} M.~Y.~Veillette and J.~T.~Chalker, Phys.~Rev.~B {\bf74}, 052402 (2006).
\bibitem{Alicea} J.~Alicea and M.~P.~A.~Fisher, Phys.~Rev.~B {\bf75}, 144411 (2007).
\bibitem{Starykh1} O.~A.~Starykh and L.~Balents, Phys.~Rev.~Lett.~{\bf98}, 077205 (2007).
\bibitem{Coldea} R.~Coldea, D.~A.~Tennant, A.~M.~Tsvelik, and Z.~Tylczynski, Phys.~Rev.~Lett.~{\bf86}, 1335 (2001).
\bibitem{Radu} T.~Radu, H.~Wilhelm, V.~Yushankhai, D.~Kovrizhin, R.~Coldea, Z.~Tylczynski, T.~L\"{u}hmann, and F.~Steglich, 
Phys.~Rev.~Lett.~{\bf95}, 127202 (2005).
\bibitem{Tokiwa} Y.~Tokiwa, T.~Radu, R.~Coldea, H.~Wilhelm, Z.~Tylczynski, and F.~Steglich, 
Phys.~Rev.~B {\bf73}, 134414 (2006).
\bibitem{Zelenak} V.~Zele\v{n}\'{a}k, A.~Orend\'{a}\v{c}ov\'{a}, I.~C\'{i}sa\v{r}ov\'{a}, J.~\v{C}ern\'{a}k, 
O.~V.~Kravchyna, J.-H.~Park, M.~Orend\'{a}\v{c}, A.~G.~Anders, A.~Feher, and M.~W.~Meisel, Inorg.~Chem.~{\bf45}, 1774 (2006). 
\bibitem{Cuccoli} A.~Cuccoli, T.~Roscilde, R. ~Vaia, and P.~Verrucchi, Phys.~Rev.~B {\bf68}, 060402(R) (2003).
\bibitem{Goddard} P.~A.~Goddard \emph{et al.}, 
New J. Phys. {\bf 10}, 083025 (2008).
\bibitem{Xiao} F.~Xiao \emph{et al.}, 
Phys.~Rev.~B {\bf 79}, 134412 (2009).
\bibitem{Sengupta} P.~Sengupta \emph{et al.}, 
Phys.~Rev.~B {\bf79}, 060409(R) (2009). 
\bibitem{Riegel} S.~Riegel and G.~Weber, J.~Phys.~E {\bf19}, 790 (1986).
\bibitem{Bernu}B.~Bernu and G.~Misguich, Phys.~Rev.~B {\bf63}, 134409 (2001).
\bibitem{Nelson} S.~Chakravarty, B.~I.~Halperin, and D.~R.~Nelson, Phys.~Rev.~Lett.~{\bf60}, 1057 (1988).
\bibitem{Chubukov} A.~V.~Chubukov, S.~Sachdev, and J.~Ye, Phys.~Rev.~B {\bf 49}, 11919 (1994).
\bibitem{Singh} P.~Sengupta, A.~W.~Sandvik, and R.~R.~P.~Singh, Phys.~Rev.~B {\bf68}, 094423 (2003).
\bibitem{Lancaster} T.~Lancaster \emph{et al.}, 
Phys.~Rev.~B {\bf75}, 094421 (2007). 
\bibitem{Kajnak} M.~Kaj\v{n}akov\'{a}, M.~Orend\'{a}\v{c}, A.~Orend\'{a}\v{c}ov\'{a}, A.~Vl\v{c}ek, J.~\v{C}ern\'{a}k, 
O.~V.~Kravchyna, A.~G.~Anders, M.~Ba{\l}anda, J.-H.~Park, A.~Feher, and M.~W.~Meisel, Phys.~Rev.~B {\bf71}, 014435 (2005).   
\bibitem{Bonner} J.~C.~Bonner and M.~E.~Fischer, Phys.~Rev.~{\bf135}, A640 (1964).
\bibitem{Klumper} A.~Kl\"{u}mper, Eur.~Phys.~J.~B {\bf5}, 677 (1997). 
\bibitem{Villain} J.~Villain and J.~M. Loveluck, J.~Phys.~(France) Lett.~{\bf38}, L77 (1977). 
\bibitem{Pires} A.~S.~T.~Pires, Phys.~Rev.~B {\bf50}, 9592 (1994). 
\bibitem{private} T.~Roscilde (private communication).
\bibitem{thesis} T.~Roscilde, PhD. Thesis, University of Pavia, 2002.
\bibitem{Pavarini} E.~Pavarini (private communication). 
\bibitem{Ramirez} A.~P.~Ramirez, B.~Hessen, and M.~Winklemann, Phys.~Rev.~Lett.~{\bf84}, 2957 (2000).
\bibitem{Nakatsuji} S.~Nakatsuji \emph{et al.}, 
Science {\bf309}, 1697 (2005).
\bibitem{Chern} C.-H.~Chern, Phys.~Rev.~B {\bf78}, 020403(R) (2008).
\bibitem{Mac} D.~E.~MacLaughlin, Y.~Nambu, S.~Nakatsuji, R.~H.~Heffner, Lei Shu, O.~O.~Bernal, and K.~Ishida, 
Phys.~Rev.~B {\bf78}, 220403(R) (2008).

\end{thebibliography}
\end{document}